\documentclass[10pt,twocolumn,twoside]{IEEEtran}

\usepackage{amsmath}
\usepackage{color}
\usepackage{cases}
\usepackage{amsfonts,amsmath,amssymb}
\usepackage{mathrsfs}
\usepackage{cite}
\usepackage{graphicx}
\usepackage{url}
\usepackage{enumerate}
\usepackage{subfigure}
\usepackage{hyperref}
\usepackage{algorithm}
\usepackage{algpseudocode}
\usepackage{amsmath,bm}
\usepackage{caption}
\usepackage{float}
\usepackage{stfloats}
\usepackage{multirow}
\usepackage{url}
\usepackage{breakurl}
\usepackage{txfonts}
\usepackage{subfigure}
\usepackage{booktabs}
\usepackage{threeparttable}
\usepackage{enumitem}

\usepackage{geometry}
\DeclareMathOperator*{\argmax}{arg\,max}

\begin{document}

\title{\huge{Alternating Diffusion Map Based Fusion of Multimodal Brain Connectivity Networks for IQ Prediction}}

\author{\mbox{Li Xiao, Julia M. Stephen, Tony W. Wilson, Vince D. Calhoun, and Yu-Ping Wang}
\thanks{L. Xiao and Y.-P. Wang are with the Department of Biomedical Engineering,
Tulane University, New Orleans, LA 70118, (e-mail: wyp@tulane.edu).}
\thanks{J. M. Stephen and V. D. Calhoun are with the Mind Research Network, Albuquerque, NM 87106. }
\thanks{T. W. Wilson is with the Department of Neurological Sciences, University of Nebraska Medical Center, Omaha, NE 68198.}
\thanks{V. D. Calhoun is also with the Department of Electrical and Computer Engineering, University of New Mexico, Albuquerque, NM 87131.}
}
\maketitle

\begin{abstract}
\textit{Objective}: To explain individual differences in development, behavior, and cognition, most previous studies focused on projecting resting-state functional MRI (fMRI) based functional connectivity (FC) data into a low-dimensional space via linear dimensionality reduction techniques, followed by executing analysis operations. However, linear dimensionality analysis techniques may fail to capture nonlinearity of brain neuroactivity. Moreover, besides resting-state FC, FC based on task fMRI can be expected to provide complementary information. Motivated by these considerations, we nonlinearly fuse resting-state and task-based FC networks (FCNs) to seek a better representation in this paper.
\textit{Methods:} We propose a framework based on alternating diffusion map (ADM), which extracts geometry-preserving low-dimensional embeddings that successfully parameterize the intrinsic variables driving the phenomenon of interest. Specifically, we first separately build resting-state and task-based FCNs by symmetric positive definite matrices using sparse inverse covariance estimation for each subject, and then utilize the ADM to fuse them in order to extract significant low-dimensional embeddings, which are used as fingerprints to identify individuals.  \textit{Results:} The proposed framework is validated on the Philadelphia Neurodevelopmental Cohort data, where we conduct extensive experimental study on resting-state and fractal $n$-back task fMRI for the classification of intelligence quotient (IQ). The fusion of resting-state and $n$-back task fMRI by the proposed framework achieves better classification accuracy than any single fMRI, and the proposed framework is shown to outperform several other data fusion methods. \textit{Conclusion and Significance:} To our knowledge, this paper is the first to demonstrate a successful extension of the ADM to fuse resting-state and task-based fMRI data for accurate prediction of IQ.
\end{abstract}


\section{Introduction}
Over the past few decades, there has been great attention to data fusion techniques and their applications in various fields; see, e.g., \cite{vv1,vv2,daf1,daf2,daf3} and references therein.
Integrating multiple datasets acquired by different sensors for a phenomenon of interest may yield more informative knowledge than any individual dataset does, because the multiple datasets can provide complementary information of the observed phenomenon from several different views.
A straightforward approach is to simply concatenate feature vectors from multiple datasets into a single feature vector. However,
such a concatenation scheme is very sensitive to the scaling of the data.
Multivariate approaches, such as canonical correlation analysis (CCA) \cite{cca}, independent component analysis (ICA) \cite{ica}, and partial least squares (PLS) \cite{pls}, have been independently developed by maximizing the correlation between the linear combinations of features from two datasets. Their penalized versions for high-dimensional settings and extensions to multiple datasets have also been proposed in \cite{cca1,cca3,pls1,cca4}. To analyze the joint information between different tasks and different brain regions in multiple functional MRI (fMRI) datasets, Calhoun and his collaborators \cite{ica2,vinc1,vinc2,vinc3,vinc4} have proposed many ICA-based multitask data fusion approaches (e.g., joint ICA and multimodal CCA+joint ICA) according to various optimization assumptions.
All the aforementioned approaches are based on linear mixture models, so they cannot optimally handle datasets that appear to have nonlinear structures and relations. To overcome this issue, many kernel based data fusion approaches have been studied in recent years \cite{kernel2,kernel3,kernel4,kernel5,kernel6,ADM1,ADM3,ADM-use}, where each dataset is individually used to construct a kernel matrix, and then the obtained kernel matrices are combined in linear or nonlinear ways to seek a unified kernel matrix that best represents all available information. A typical kernel based approach is multiple kernel learning \cite{kernel2,kernel3}, which finds the unified kernel matrix by linearly combining the multiple kernel matrices. However, this approach assumes that the complementary information from multiple data sources is linearly provided, which might not necessarily be true in practice. Moreover, to learn an optimal unified kernel matrix, tuning the weight coefficient assigned to each single kernel matrix is computationally intensive. In \cite{kernel4,kernel5,kernel6,ADM1,ADM3,ADM-use}, nonlinear kernel fusion processes have been proposed, which represent complementary information nonlinearly into the intrinsic low-dimensional geometry, and avoid assigning weight coefficient to each single kernel matrix.

An approach of particular interest in this paper is alternating diffusion map (ADM), which was proposed more recently in \cite{ADM1,ADM3,ADM-use}.
The ADM is based on the framework of diffusion map (DM) \cite{DM1}, one class of manifold learning algorithms \cite{egp}, and can achieve nonlinear dimensionality reduction in such a way that the intrinsic common structures underlying multiple high-dimensional datasets are maintained.
More concretely, the ADM takes advantage of the product of the kernel matrices constructed separately by each dataset based on a stochastic Markov matrix to produce a unified representation, which can be interpreted as employing diffusion processes on each dataset in an alternating manner. This allows one to extract the common latent variables across multiple datasets that are assumed to drive the observed phenomenon, while filtering out other hidden variables that are sensor-specific and thought of as nuisance, irrelevant to the phenomenon. Hence, the ADM can provide a more reliable description of the phenomenon. 
So far the ADM has proven to be a powerful tool in voice detection from audio-visual signals \cite{amdapp2,amdapp4}, Alzheimer's disease classification from multiple electroencephalography (EEG) signals \cite{amdapp5}, and sleep stage classification from EEG and respiration signals \cite{amdapp6}. Here we show that the ADM can also be adapted to multimodal fMRI data. To our knowledge, this paper is the first to demonstrate a successful extension of the ADM to fuse resting-state and task-based fMRI data for the prediction of intelligence quotient (IQ).

The proposed framework in this paper begins with a preprocessing stage in which a brain functional connectivity network (FCN) is individually built for each subject from fMRI data. More specifically, the brain is graphically depicted as a network with regions of interest (ROIs) as the nodes and functional connectivities (FCs) as the edges, where the FC between two nodes is defined as statistical dependence between the blood oxygenation level-dependent (BOLD) fMRI time series in the two corresponding ROIs.
Different from conventionally representing the FCN by a sample covariance matrix of the multi-ROI time series, we represent it by a symmetric positive definite (SPD) matrix \cite{logecu2,logecu3,logecu4}, which is computed based on sparse inverse covariance estimation using the graphical least absolute shrinkage and selection operator (GLASSO) algorithm \cite{glass}. Accordingly, two sets of SPD matrices are respectively derived from resting-state and task-based fMRI datasets. The FCN organization varies between individuals, and accordingly acts as a ``fingerprint'' of a subject \cite{amico2}. Recent works \cite{strucpp1,strucpp2,strucpp3} have also studied the relations between the functional and structural brain connectivity patterns to improve the reliability of individual ``fingerprint'' as a biomarker.

We therefore store the SPD matrices of all subjects and treat them as new features from fMRI data for subsequent analysis. However, the dimension of the SPD matrix is usually much larger than the number of subjects.
For example, there are $34716$ FCs with $264$ ROIs in our study.
If we directly use the SPD matrices to train a classifier, it will suffer from the curse of dimensionality, which often leads to overfitting and poor generalization performance. Fortunately, despite individual variation, human brains do in fact share common connectivity patterns across different subjects, i.e., variations of the SPD matrices representing brain connectivity are driven by a small subset of unknown parameters. It suggests that we adopt nonlinear dimensionality reduction (e.g., manifold learning) algorithms to extract the intrinsic variables of the SPD matrices prior to training a classifier. In this paper, based on the two sets of SPD matrices derived from two fMRI datasets, respectively, we use the ADM to fuse them to find meaningful low-dimensional embeddings, so that their shared source of variability is maintained while noise specific to any single set of SPD matrices is reduced. These low-dimensional embeddings are then used as fingerprints to classify individuals of different behaviors and cognitions (e.g., IQ).

As the set of SPD matrices is known to form a Riemannian manifold instead of a full Euclidean space, geometric distances, such as affine-invariant Riemannian distance \cite{airm} and root stein divergence distance \cite{rsm}), have been proposed to measure the similarities of SPD matrices by considering the underlying manifold where they reside. These distances can better discover the Riemannian geometry than the traditional Euclidean distance, and have been used successfully to characterize FC differences \cite{logecu2,logecu3,logecu4,logecu1}.
In this paper, we adopt a geodesic distance on SPD matrices, namely the Log-Euclidean distance \cite{leudis}, to measure the similarities of SPD matrices in the ADM because of its computational efficiency. The Euclidean distance and the Cholesky distance \cite{ckdis} are tested for comparison.

We finally validate our proposed framework by fusing two fMRI datasets (i.e., resting-state and fractal $n$-back task fMRI) from the publicly available Philadelphia Neurodevelopmental Cohort (PNC) data \cite{pncdata1,pncdata2} to build a predictor for subjects with different IQs. The subjects' IQ scores were assessed by the Wide Range Achievement Test (WRAT) administered in the PNC. The WRAT is an achievement test that measures an individual's learning ability including reading, spelling, and arithmetic \cite{robert}, which can provide a reliable estimate of IQ. A large body of clinical studies has emerged to argue that distinct patterns of brain functional activity account for the proportion of difference of IQ among individuals \cite{human1,human3}. These findings suggest that the ADM of fusing multiple sets of FCNs in this paper has the potential to automate the task of classifying populations of low and high IQs.
As will be seen experimentally, the classification results demonstrate the advantage of our proposed framework. Specifically, the ADM achieves superior classification performance over that of the DM  (using any single set of FCNs) and several existing fusion methods. In addition, the effectiveness of incorporating the log-Euclidean distance into the DM and the ADM is verified in comparison to the Euclidean and Cholesky distances.

The rest of this paper is organized as follows. In Section \ref{methodmaterial}, the proposed framework is presented, including the brain FCN construction and two manifold learning methods (i.e., the DM and the ADM). In Section \ref{resultsper}, a simulation example is first illustrated, and then the experimental results on the PNC data are shown. The discussions are presented in Section \ref{future}, followed by the conclusion in Section \ref{condul}.

\textit{Notations:} Uppercase boldface, lowercase boldface, and normal italic letters are used to denote matrices, vectors, and scalars, respectively. The superscript $T$ denotes the transpose of a vector or matrix.
$A_{i,j}$ denotes the $(i,j)$-th entry of matrix $\bm{A}$, and $a(i)$ denotes the $i$-th entry of vector $\bm{a}$. We denote the set of real numbers as $\mathbb{R}$.

\section{Methods}\label{methodmaterial}
\begin{figure*}[!t]
  \centering
  \includegraphics[width=1.87\columnwidth]{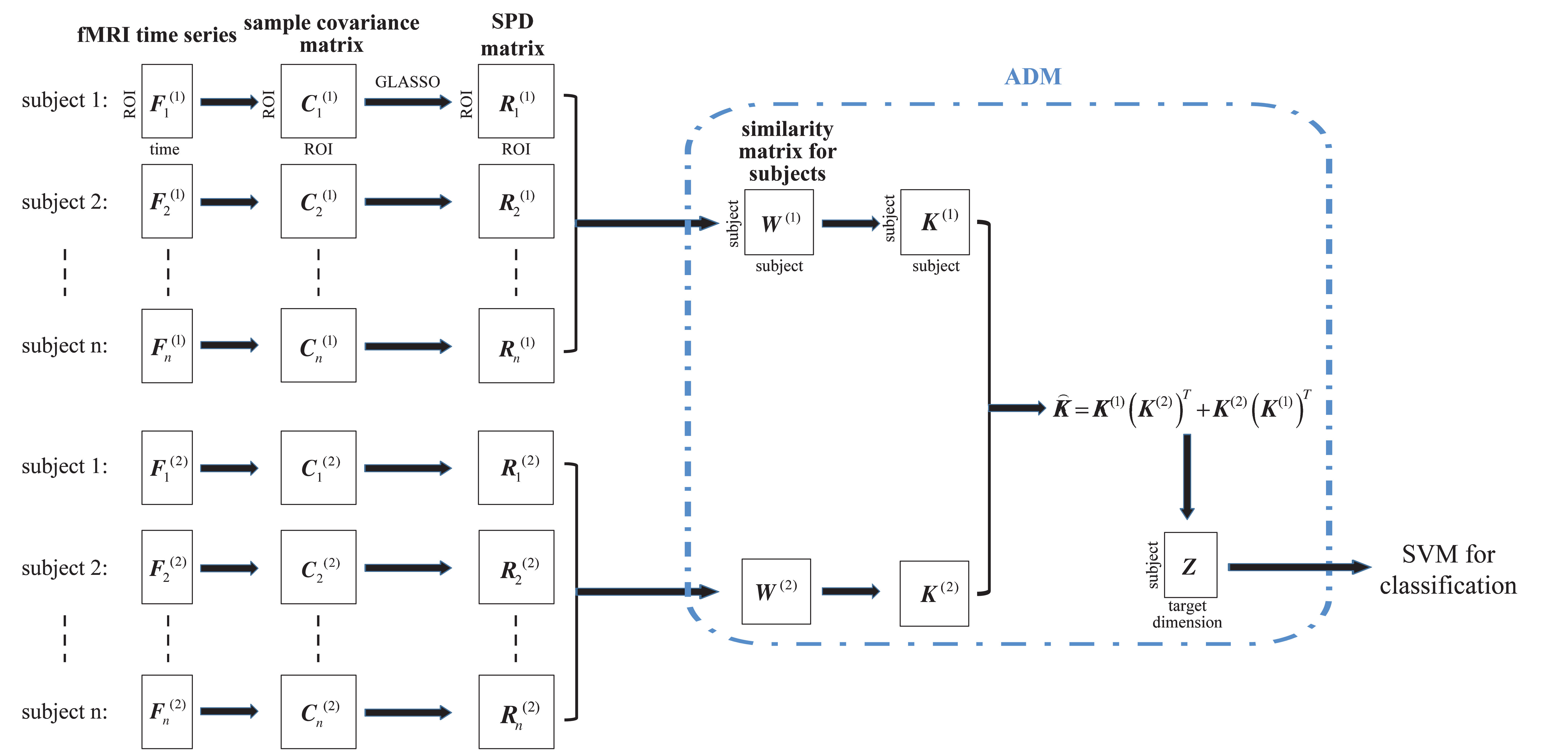}\\
  \caption{The flowchart of the proposed framework. $\left\{\bm{F}_i^{(1)}\right\}_{i=1}^{n}$ and $\left\{\bm{F}_i^{(2)}\right\}_{i=1}^{n}$ denote two fMRI datasets for the same $n$ subjects. $\bm{R}_{i}^{(j)}$ is a SPD matrix that represents the brain FCN of subject $i$ based on the $j$-th fMRI dataset. The ADM is applied to the fusion of brain FCNs derived from the two fMRI datasets in order to obtain a low-dimensional representation $\bm{Z}$.}\label{fig1}
  \vspace{-0.1cm}
\end{figure*}

In this section, an overview of the proposed framework is outlined in Fig. \ref{fig1}. There are three major steps: $(1)$ brain FCNs are extracted as SPD matrices from each fMRI dataset; $(2)$ the ADM is applied for fusing two sets of FCNs derived from two fMRI datasets to find a meaningful low-dimensional representation; and $(3)$ support vector machine (SVM) classification is carried out based on the low-dimensional embeddings. In the following, we will present the details of these steps.

\subsection{Brain FCN representation using SPD matrices}\label{brain}
The BOLD fMRI signal, as a time series, measures neural activity by detecting changes in blood flow at many spatial locations of the brain. In fMRI, studies can focus on specific tasks as well as at rest, and brain networks are usually built based on the BOLD signals to describe FC across brain regions. The network nodes are brain ROIs, and the FC between two nodes is defined as temporal covariance or correlation of fMRI time series in the two nodes.

Let $\bm{F}=\left[\bm{f}_1,\bm{f}_2,\cdots,\bm{f}_m\right]\in\mathbb{R}^{p\times m}$ be a BOLD fMRI time series for a subject, where $m$ is the number of time points and  $\bm{f}_i\in\mathbb{R}^p$ is a $p$-dimensional vector, corresponding to an observation of $p$ brain ROIs at the $i$-th time point. Assume that $\bm{F}$ has been normalized to have zero mean and unit variance along each row. As described above, the FCN is represented by a covariance matrix $\bm{R}$ of the multi-ROI time series.
To estimate $\bm{R}$, we generally obtain the estimation of its inverse $\bm{S}=\bm{R}^{-1}$ by maximizing the penalized log-likelihood over the space of all $p\times p$ SPD matrices:
\begin{equation}\label{maxi}
\argmax_{\bm{S}\succ0}\log(\det(\bm{S}))-\mbox{tr}(\bm{C}\bm{S})-\lambda\lVert\bm{S}\rVert_1,
\end{equation}
where $\bm{C}=\frac{1}{m}\bm{F}\bm{F}^T\in\mathbb{R}^{p\times p}$ is the sample covariance matrix, and $\det(\cdot)$, $\mbox{tr}(\cdot)$, $\lVert\cdot\rVert_1$ denote the determinant, the trace, the sum of the absolute values of the entries of a matrix, respectively. In (\ref{maxi}),
the regularization parameter $\lambda>0$ controls the tradeoff between the degree of sparsity and the log-likelihood estimation of $\bm{S}$. In this paper, we use the Bayesian Information Criterion (BIC) \cite{schwaz} to select the optimum $\lambda$, and
the maximization problem (\ref{maxi}) can be efficiently solved via the graphical LASSO (GLASSO) algorithm \cite{glass} (its Matlab software package: \url{http://statweb.stanford.edu/~tibs/glasso/}).

\subsection{Nonlinear dimensionality reduction of FCNs}\label{manifold}
From (\ref{maxi}), we can individually compute the SPD matrices $\bm{R}_i$, $i=1,2,\cdots,n$, to represent the FCNs of $n$ subjects from one fMRI dataset. In what follows, we shall use the terms ``SPD matrices'' and ``FCNs'' interchangeably. The SPD matrices are treated as the features extracted from subjects' fMRI data for subsequent analysis, and considered as points distributed in a high-dimensional space. We have to reduce the dimension of these SPD matrices by finding the significant features, since many of the features may be noninformative or redundant while increasing computational cost and classification complexity.
In spite of individual variation, brains do share some common FC patterns across different subjects. Therefore, the SPD matrices used to represent FCNs shall have some similar structures \cite{pascal}, and their variations only depend on a small subset of unknown parameters. Inspired by this evidence, we aim to generate a low-dimensional representation of the SPD matrices. Since brain activity involves multiple nonlinear neural dynamics, we adopt here a nonlinear dimensionality reduction algorithm for best representing the high-dimensional SPD matrices by their low-dimensional embeddings, where the intrinsic geometry of the SPD matrices can be well preserved in the embedding coordinates. The details are elaborated as follows.

\subsubsection{Gaussian kernel function}
In machine learning, kernel functions are often used to define similarity measures to learn the relations among subjects via the kernel trick, and in particular the Gaussian kernels are widely used.
In this paper, we calculate a similarity matrix by using the Gaussian kernel function with a distance of SPD matrices, i.e.,
\begin{equation}\label{weightmatrix}
     W_{i,j}=\exp\left(-\frac{{\rm{d}^2}(\bm{R}_i,\bm{R}_j)}{\sigma}\right),
\end{equation}
where $\sigma>0$ is the bandwidth of the Gaussian kernel function and ${\rm{d}}(\cdot,\cdot)$ is a distance chosen by the user to measure two SPD matrices. This construction defines
a weighted graph, in which the nodes correspond to the $n$ subjects $\left\{\bm{R}_i\right\}_{i=1}^{n}$, and $\bm{W}\in\mathbb{R}^{n\times n}$ is the weight matrix of the graph.

Different definitions of ${\rm{d}}(\cdot,\cdot)$ would lead to different similarity matrices. An appropriate distance is crucial to perform the following dimension reduction while revealing the intrinsic geometry of the SPD matrices, since the set of SPD matrices is restricted to some Riemannian manifold, not a full Euclidean space. For ease of computation, we investigate one commonly used geodesic distance, i.e., the log-Euclidean distance (LEU) \cite{leudis}, that considers the specific geometry of the manifold.
The LEU between $\bm{R}_i$ and $\bm{R}_j$ is given by
\begin{equation}\label{logeuc}
{\rm{d}_{leu}}(\bm{R}_i,\bm{R}_j)=\lVert\log(\bm{R}_i)-\log(\bm{R}_j)\rVert_{F},
\end{equation}
where, for a SPD matrix $\bm{R}\in\mathbb{R}^{p\times p}$ with its eigenvalue decomposition $\bm{R}=\bm{U}\cdot\mbox{Diag}(\mu_1,\cdots,\mu_p)\cdot\bm{U}^{T}$, the matrix logarithm of $\bm{R}$ is defined by $\log(\bm{A})=\bm{U}\cdot\mbox{Diag}(\log(\mu_1),\cdots,\log(\mu_p))\cdot\bm{U}^{T}$, and $\lVert\cdot\rVert_{F}$ denotes the Frobenius matrix norm. For comparison, we consider
the Cholesky distance (CK) \cite{ckdis} and the traditional Euclidean distance (EU) as well.
The CK is given by
\begin{equation}\label{chock}
{\rm{d}_{ck}}(\bm{R}_i,\bm{R}_j)=\lVert(\bm{R}_i)_{low}-(\bm{R}_j)_{low}\rVert_{F},
\end{equation}
where $\bm{R}_{low}$ denotes the low triangular matrix obtained by the Cholesky decomposition of $\bm{R}$, i.e., $\bm{R}=(\bm{R})_{low}(\bm{R})^{T}_{low}$.  The EU is given by
\begin{equation}\label{euclidean}
{\rm{d}_{eu}}(\bm{R}_i,\bm{R}_j)=\lVert\bm{R}_i-\bm{R}_j\rVert_{F}.
\end{equation}

The LEU (\ref{logeuc}) is known as one of the most widely adopted distances for SPD matrices, because it is a geodesic distance induced by Riemannian metrics and provides a more accurate distance measure than the EU (\ref{euclidean}). Apart from these geodesic distances, a number of other distances (e.g., the CK (\ref{chock})) that do not necessarily arise from Riemannian metrics can also be used to capture the nonlinearity among SPD matrices. Different from the LEU that is derived based on matrix logarithm, the CK induces a reparameterization measure of SPD matrices based on matrix decomposition, because a SPD matrix has a unique Cholesky decomposition.
It is shown in \cite{ckdis} that the Gaussian kernels (\ref{weightmatrix}) with the LEU, the CK, and the EU are all positive semidefinite on manifolds for any $\sigma>0$, such that one would be able to freely tune $\sigma$ to reflect the data distribution.

\subsubsection{DM for single FCN dataset}\label{dm}
\begin{figure}[!t]
 \centering
  \includegraphics[width=0.86\columnwidth]{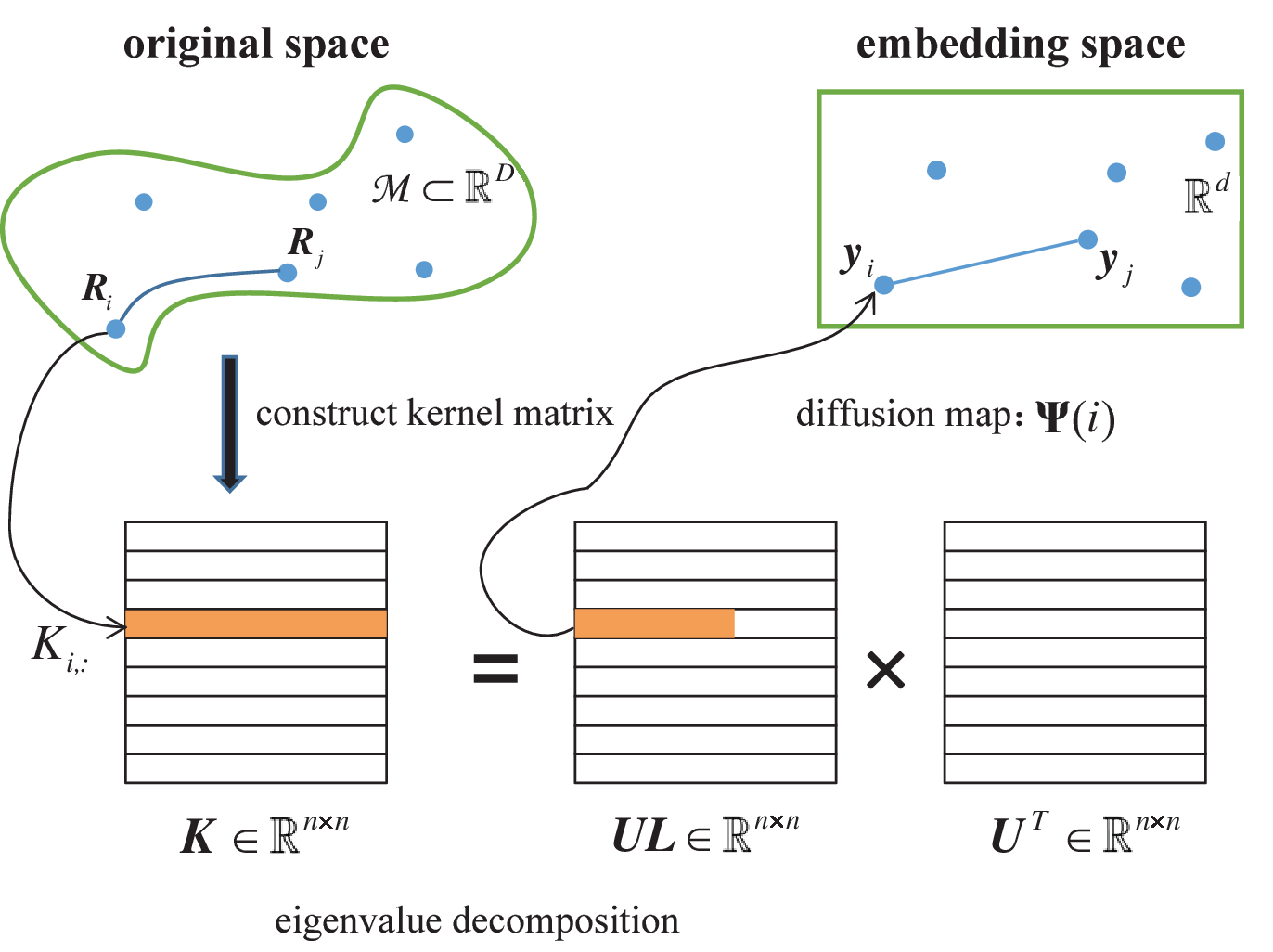}\\
  \caption{The DM for a single dataset $\left\{\bm{R}_i\right\}_{i=1}^{n}$, where $\bm{K}$ is a normalized kernel matrix (\ref{kernelcom}), and $\bm{U},\bm{L}$ are its the eigendecomposition components (\ref{eigendecom}). Assume that the high-dimensional data points $\left\{\bm{R}_i\right\}_{i=1}^{n}$ approximately lie on a low-dimensional manifold $\mathcal{M}$ embedded into $\mathbb{R}^{D}$. With the DM, $\left\{\bm{R}_i\right\}_{i=1}^{n}$ are mapped into the geometry-preserving low-dimensional embeddings $\left\{\bm{y}_i\right\}_{i=1}^{n}\in\mathbb{R}^{d}$.}\label{fig2}
  \vspace{-0.1cm}
\end{figure}

Considering that the data points $\left\{\bm{R}_i\right\}_{i=1}^{n}$ lie on an intrinsically low-dimensional manifold embedded into $\mathbb{R}^{D}$, we use the DM \cite{DM1} to obtain their low-dimensional embeddings $\left\{\bm{y}_i\right\}_{i=1}^{n}\in\mathbb{R}^{d}$ with $d\ll D$. The DM is a graph-based nonlinear dimensionality reduction method, which extends and enhances ideas from other manifold learning methods by deploying a stochastic Markov matrix based on the similarities between data points in high-dimensional space to identify a low-dimensional representation that captures the intrinsic geometry in the dataset. The procedure of the DM is demonstrated in Fig. \ref{fig2} and described in detail below.

Based on the similarity matrix $\bm{W}$ calculated in (\ref{weightmatrix}), we first get a normalized kernel matrix $\bm{K}$ by
\begin{equation}\label{kernelcom}
\bm{K}=\bm{Q}^{-1}\bm{W}
\end{equation}
such that each row sums to $1$,
where $\bm{Q}\in\mathbb{R}^{n\times n}$ is a diagonal matrix with $Q_{l,l}=\sum_{j=1}^{n}W_{l,j}$. Hence, we can imagine a Markov chain on the graph with the transition matrix $\bm{K}$, in the sense that the $(i,j)$-th entry $K_{i,j}$ represents the transition probability from node $i$ to node $j$.

It is easy to check that $\bm{K}$ is similar to the positive semidefinite matrix $\bm{Q}^{-1/2}\bm{W}\bm{Q}^{-1/2}$. As such, let $\lambda_0\geq\lambda_1\geq\cdots\geq\lambda_{n-1}\geq0$ and $\bm{\psi}_0,\bm{\psi}_1,\cdots,\bm{\psi}_{n-1}$ denote the ordered eigenvalues and corresponding normalized eigenvectors of $\bm{K}$, i.e.,
\begin{equation}\label{eigendecom}
\bm{K}=\bm{U}\bm{L}\bm{U}^{T},
\end{equation}
where $\bm{U}=[\bm{\psi}_0,\bm{\psi}_1,\cdots,\bm{\psi}_{n-1}]$ and $\bm{L}=\mbox{diag}(\lambda_0,\lambda_1,\cdots,\lambda_{n-1})$. Moreover, we can readily verify that the largest eigenvalue $\lambda_0$ is equal to $1$ and its associated eigenvector $\bm{\psi}_0$ is a constant vector.
Then, a compact representation, referred to as DM, is achieved by keeping only the $d$ largest non-trivial eigenvalues and eigenvectors of $\bm{K}$, i.e.,
\begin{equation}
\Psi(i):\; \bm{R}_i\longmapsto\bm{y}_i\triangleq\left[\lambda_1\psi_1(i),\lambda_2\psi_2(i),\cdots,\lambda_{d}\psi_{d}(i)\right]^{T},
\end{equation}
where $d$ is an estimated dimension of the embedding space.

The key idea in the DM is that the Euclidean distance between two embeddings (e.g., $\bm{y}_i$ and $\bm{y}_j$) is approximately equal to the diffusion distance between the two corresponding data points (e.g., $\bm{R}_i$ and $\bm{R}_j$) in the original space. The diffusion distance between the $i$-th and $j$-th subjects is defined as the weighted $L_2$ distance between the transition probabilities of node $i$ and node $j$, i.e.,
\begin{equation}\label{defindif}
\mathcal{D}(i,j)=\sqrt{\sum_{l=1}^n\frac{(K_{i,l}-K_{j,l})^2}{\phi(l)}},
\end{equation}
where $\bm{\phi}$ stands for the stationary distribution of $\bm{K}$, calculated by $\phi(l)=Q_{l,l}/\sum_{i=1}^{n}Q_{i,i}$ for $1\leq l\leq n$. The diffusion distance is a metric that can reveal the intrinsic geometry among data points. It is robust to noise as well, since the diffusion could be viewed as a nonlinear process that averages all possible connectivity between pairs of data points on the graph.

\subsubsection{ADM based fusion of two FCN datasets}
The ADM \cite{ADM1,ADM3,ADM-use} is a recently developed data fusion technique on the basis of the DM framework. The purpose of the ADM is to fuse two datasets to find a more coherent and accurate representation, in the sense that the information from the two datasets is diffused to yield the underlying common information (which is assumed to drive the phenomenon of interest), and meanwhile nuisance specific to any single dataset is reduced.
Let $\left\{\bm{R}^{(1)}_i\right\}_{i=1}^{n}$ and $\left\{\bm{R}^{(2)}_i\right\}_{i=1}^{n}$ be the FCNs extracted from two different fMRI datasets for the same $n$ subjects, respectively. By using the ADM described below, we can obtain low-dimensional embeddings $\left\{\bm{z}_i\right\}_{i=1}^{n}\in\mathbb{R}^{\tilde{d}}$.

In the same way as in (\ref{weightmatrix}), we separately construct similarity matrices of the two datasets: for all $1\leq i,j\leq n$ and $l=1,2$,
\begin{align}\label{gauker}
W^{(l)}_{i,j}=\exp\left(-\frac{{\rm{d}}^2(\bm{R}^{(l)}_i,\bm{R}^{(l)}_j)}{\sigma_l}\right),
\end{align}
where $\sigma_l$ is the tuneable kernel bandwidth and ${\rm{d}}(\cdot,\cdot)$ is a chosen metric on the data points. From the similarity matrices, we get the normalized kernel matrices $\bm{K}^{(1)}$ and $\bm{K}^{(2)}$ as in (\ref{kernelcom}), respectively. According to the ADM in \cite{ADM-use}, a unified kernel matrix is given by
\begin{equation}\label{unif}
\widehat{\bm{K}}=\bm{K}^{(1)}\left(\bm{K}^{(2)}\right)^{T}+\bm{K}^{(2)}\left(\bm{K}^{(1)}\right)^{T}.
\end{equation}
Since $\widehat{\bm{K}}$ is real and symmetric, it has real eigenvalues, and the eigenvectors are real and orthogonal to each other. As such, let $|\widetilde{\lambda}_0|\geq|\widetilde{\lambda}_1|\geq\cdots\geq|\widetilde{\lambda}_{n-1}|$ be the eigenvalues of $\widehat{\bm{K}}$ with decreasing magnitude, and $\widetilde{\bm{\psi}}_0,\widetilde{\bm{\psi}}_1,\cdots,\widetilde{\bm{\psi}}_{n-1}$ be the corresponding normalized eigenvectors.
Hence, a low-dimensional representation (referred to as ADM) for the common structures in the datasets is obtained by taking its eigenvectors corresponding to the $\tilde{d}$ largest eigenvalues in magnitude, i.e.,
\begin{equation}\label{effi}
\widetilde{\Psi}(i):\; (\bm{R}^{(1)}_i,\bm{R}^{(2)}_i)\longmapsto\bm{z}_i
\triangleq\left[\widetilde{\psi}_0(i),\widetilde{\psi}_1(i),\cdots,\widetilde{\psi}_{\tilde{d}-1}(i)\right]^{T},
\end{equation}
where $\tilde{d}$ is an estimated dimension of the embedding space.

In the ADM, a Markov chain on a graph is first built for each dataset, where the subjects represent the graph nodes, and the normalized kernel matrix is viewed as the transition matrix of the Markov chain on the graph. In other words, we obtain two graphs with the same set of nodes and two different transition matrices (i.e., $\bm{K}^{(1)}$ and $\bm{K}^{(2)}$). Then, we combine the information from the two datasets by the product of the transition matrices,
which takes into account all the various connectivities of two nodes hopping within and across the two graphs.
It is shown in \cite{ADM-use} that efficient low-dimensional embeddings (\ref{effi}) based on the matrix $\widehat{\bm{K}}$ characterize the common structures (common latent variables) between the manifolds underlying the different datasets, and in the meantime attenuate the differences (sensor-specific variables) between the manifolds. The interested reader can find a theoretical foundation of the ADM in \cite{ADM1,ADM-use}.

\subsubsection{Out-of-sample extension}
In the above, we present how to use the DM (or ADM) to provide a mapping for a training set with $n$ FCNs $\left\{\bm{R}_i\right\}_{i=1}^{n}$ (or $2n$ FCNs $\left\{\bm{R}^{(1)}_i\right\}_{i=1}^{n}$ and $\left\{\bm{R}^{(2)}_i\right\}_{i=1}^{n}$) to a $d$-dimensional (or $\tilde{d}$-dimensional) space. In order to extend the mapping to new data points (unlabeled FCNs) without reapplying a large-scale eigendecomposition on the entire data, we introduce the Nystr\"{o}m extension \cite{nystrom,music,logecu2,kernel6}, which is an efficient non-parametric solution widely used for the methods involving spectral decomposition. Accordingly, for the DM and the ADM, respectively, we derive an explicit mapping between new FCNs and the low-dimensional embedding space obtained from the training set as follows.

Given a new FCN $\bm{R}_{n+1}$, we want to extend the DM mapping to get $\bm{y}_{n+1}$. We first calculate the similarities $W_{n+1,j}$ between $\bm{R}_{n+1}$ and $\bm{R}_{j}$, $j=1,2,\cdots,n$, and then normalize them to get $K_{n+1,j}$ for $1\leq j\leq n$, i.e.,
\begin{equation}\label{houll}
W_{n+1,j}=\exp\left(-\frac{{\rm{d}^2}(\bm{R}_{n+1},\bm{R}_j)}{\sigma}\right),\; K_{n+1,j}=\frac{W_{n+1,j}}{\sum_{i=1}^{n}W_{n+1,i}}.
\end{equation}
The extended eigenvectors for the new data point are approximated as the weighted sums of the original eigenvectors, i.e.,
\begin{equation}
\overline{\psi}_{i}(n+1)=\frac{1}{\lambda_i}\sum_{j=1}^{n}K_{n+1,j}\,\psi_{i}(j),
\end{equation}
and the embedding $\bm{y}_{n+1}$ is given by
\begin{equation}
\bm{y}_{n+1}\triangleq\left[\lambda_1\overline{\psi}_{1}(n+1),\lambda_2\overline{\psi}_{2}(n+1),\cdots,\lambda_d\overline{\psi}_{d}(n+1)\right]^{T}\in\mathbb{R}^{d}.
\end{equation}

Given new FCNs $\bm{R}^{(1)}_{n+1}$ and $\bm{R}^{(2)}_{n+1}$ for a two-dataset scenario, we want to extend the ADM mapping to get $\bm{z}_{n+1}$. Similar to (\ref{houll}), we calculate, for $1\leq j\leq n$ and $l=1,2$,
\begin{equation}
W^{(l)}_{n+1,j}=\exp\left(-\frac{{\rm{d}}^2(\bm{R}^{(l)}_{n+1},\bm{R}^{(l)}_j)}{\sigma_l}\right),\;K^{(l)}_{n+1,j}=\frac{W^{(l)}_{n+1,j}}{\sum_{i=1}^{n}W^{(l)}_{n+1,i}}.
\end{equation}
Let $\bm{K}^{(l)}_{n+1}\triangleq\left[K^{(l)}_{n+1,1},K^{(l)}_{n+1,2},\cdots,K^{(l)}_{n+1,n}\right]\in\mathbb{R}^{n}$ for $l=1,2$, and
\begin{equation}
\widehat{\bm{K}}_{n+1}\triangleq\bm{K}^{(1)}_{n+1}\left(\bm{K}^{(2)}\right)^{T}+\bm{K}^{(2)}_{n+1}\left(\bm{K}^{(1)}\right)^{T}\in\mathbb{R}^{n}.
\end{equation}
Then, the extension is given by
\begin{equation}
\overline{\widetilde{\psi}}_i(n+1)=\frac{1}{\widetilde{\lambda}_i}\sum_{j=1}^{n}\widehat{\bm{K}}_{n+1}(j)\,\widetilde{\psi}_{i}(j),
\end{equation}
and the embedding $\bm{z}_{n+1}$ is
\begin{equation}
\bm{z}_{n+1}\triangleq\left[\overline{\widetilde{\psi}}_0(n+1),\overline{\widetilde{\psi}}_1(n+1),\cdots,\overline{\widetilde{\psi}}_{\tilde{d}-1}(n+1)\right]^{T}\in\mathbb{R}^{\tilde{d}}.
\end{equation}

\subsection{Classification using SVM}\label{papr}
In this paper, classification is explored as a potential application to validate our proposed framework, in that if the intrinsic manifold structures of data are faithfully preserved by the proposed framework, the obtained embeddings of the original high-dimensional data points that belong to different classes will be separated far from each other in the low-dimensional embedding space. The classification performance is assessed by
using a linear kernel SVM with default hyper-parameters on the embeddings.
We remark that we here choose a simple linear kernel SVM classifier for three reasons: 1) since the DM and the ADM mentioned above provide embedded features globally in linear coordinates, we limited the tests to linear classifiers; 2) SVM is known as one of the state-of-the-art classifiers and has been extensively used in biomedical data analysis because of its accurate classification performance \cite{svm1,svm3}; and 3)
although there are many other advanced classifiers, the emphasis in this paper is the superior performance of the proposed framework, not the optimal classification scheme.

\section{Experimental results and discussion}\label{resultsper}
\subsection{Simulation result}
\begin{figure}[!h]
  \centering
  \hspace*{-0.9cm}\includegraphics[width=1.2\columnwidth]{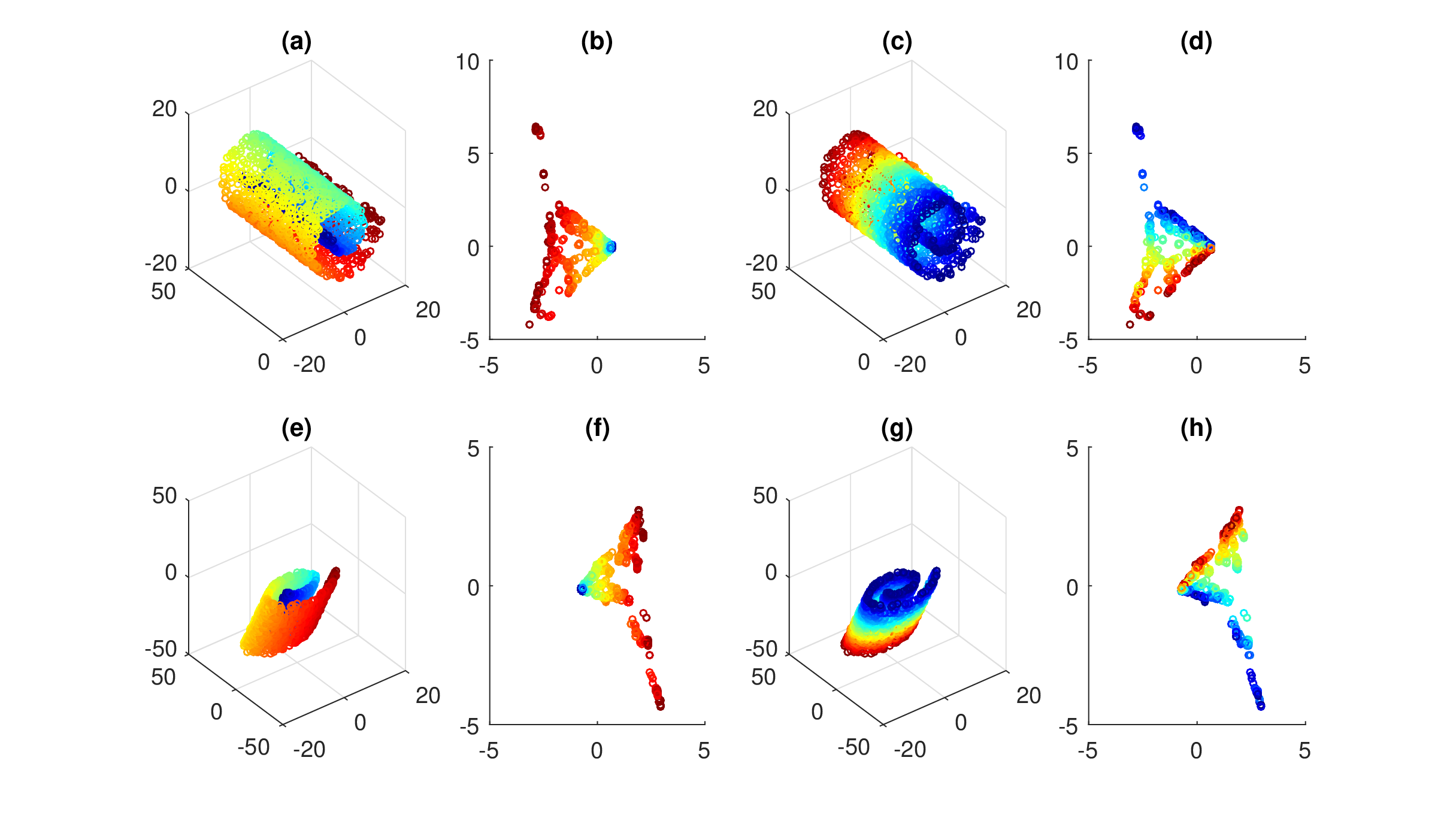}\\
  \caption{The scatter plots of two $3$-dimensional Swiss roll datasets and their $2$-dimensional embeddings. Top row: subfigures obtained from $\left\{\bm{s}_i^{(1)}\right\}_{i=1}^{n}$. Bottom row: subfigures obtained from $\left\{\bm{s}_i^{(2)}\right\}_{i=1}^{n}$. For example, (a), (c) are scatter plots of $\left\{\bm{s}_i^{(1)}\right\}_{i=1}^{n}$, and (b), (d) are their $2$-dimensional embeddings. Points in the first two columns are colored according to the common variable $\theta$, and those in the last two columns are colored according to their own sensor-specific variables.}\label{dmd}
  \vspace{-0.1cm}
\end{figure}

Let $x,y,\theta$ be three statistically independent uniform random variables on $(0,1)$. We generate $n=2000$ samples $(x_i,y_i,\theta_i)$ of $(x,y,\theta)$, and define two sets of simulated samples in $\mathbb{R}^3$ by
\begin{equation}
\bm{s}_i^{(1)}=\left[
                 \begin{array}{c}
                   \pi(1+3\theta_i)\cos(\pi(1+3\theta_i)) \\
                   50x_i \\
                    \pi(1+3\theta_i)\sin(\pi(1+3\theta_i))\\
                 \end{array}
               \right]\nonumber
\end{equation}
and
\begin{equation}
\bm{s}_i^{(2)}=\underbrace{\left[
                 \begin{array}{ccc}
                   1 & 0 & 0 \\
                   0 & 0.5 & \sqrt{3}/2 \\
                   0 & -\sqrt{3}/2 & 0.5 \\
                 \end{array}
               \right]}_{\bm{\Gamma}}
\left[
                 \begin{array}{c}
                   \pi(1+3\theta_i)\cos(\pi(1+3\theta_i)) \\
                   50y_i \\
                    \pi(1+3\theta_i)\sin(\pi(1+3\theta_i))\\
                 \end{array}
               \right]\nonumber
\end{equation}
for $1\leq i\leq n$, where $\bm{\Gamma}$ is an orthonormal transformation matrix. Assume that these two datasets are observations acquired by two sensors, respectively, where $\theta$ is a common variable, and $x$ and $y$ are the variables that are sensor-specific.
As can be seen in the first and third columns of Fig. \ref{dmd}, each set of simulated samples lies on a $2$-dimensional Swiss roll embedded in $\mathbb{R}^3$.

We first apply the DM separately to each dataset, and the $2$-dimensional embeddings are presented in the 2nd and 4th columns of Fig. \ref{dmd}. The subfigures in each row are obtained from the same dataset, i.e., (a)--(d) are scatter plots of $\left\{\bm{s}_i^{(1)}\right\}_{i=1}^{n}$ and their embeddings, and (e)--(h) are scatter plots of $\left\{\bm{s}_i^{(2)}\right\}_{i=1}^{n}$ and their embeddings. In the first two columns of Fig. \ref{dmd}, data points are colored according to $\theta_i$. In subfigures (c), (d), data points are colored according to $x_i$. In subfigures (g), (h), data points are colored according to $y_i$. One can see that all the scatter plots of the $2$-dimensional embeddings exhibit a smooth color gradient, which implies accurate parametrization of both the common and the sensor-specific variables for each dataset.

\begin{figure}[H]
  \centering
  \hspace*{-0.4cm}\includegraphics[width=1.1\columnwidth]{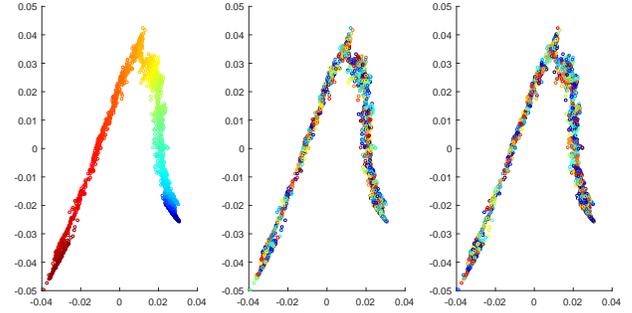}\\
  \caption{The scatter plots of $2$-dimensional embeddings obtained by the ADM on the two datasets. Points in the subfigures (from left to right) are respectively colored according to the common variable $\theta$ and sensor-specific variables $x,y$.}\label{admd}
  \vspace{-0.1cm}
\end{figure}

We next apply the ADM to fuse the two datasets. The $2$-dimensional embeddings are shown with different color coding schemes in Fig. \ref{admd}. The data points in the leftmost subfigure are colored according to the common variable $\theta$, while those in the middle and the rightmost subfigures are colored according to the sensor-specific variables $x$ and $y$, respectively. We observe that the color gradient is smooth only for the common variable. Equivalently, this means that the embeddings obtained by the ADM successfully extract a parametrization of the common variable $\theta$, while filtering out the nuisance variables $x$ and $y$ that are specific to each dataset.

\subsection{Application to IQ classification}
\subsubsection{Data preprocessing and experimental setting}
The PNC \cite{pncdata1,pncdata2} is a large-scale collaborative study of child development between the Children's Hospital of Philadelphia and the Brain Behavior Laboratory at the University of Pennsylvania. The publicly available PNC data were downloaded from dbGap (\url{www.ncbi.nlm.nih.gov/projects/gap/cgi-bin/study.cgi?study_id=phs000607.v1.p1}).
In this PNC sample, genetics, neuroimaging, and cognitive assessment measures were all acquired in nearly 900 adolescents aged from $8$ to $21$ years. In this paper, we study two functional imaging datasets (i.e., functional imaging of working memory task and resting state), and their classification performance on IQ. The scores of the WRAT administered in the PNC reflect subjects' IQ levels, since the WRAT is a standardized achievement test that measures an individual's ability, e.g., reading recognition, spelling, and math computation \cite{robert}, which can provide a reliable estimate of IQ. To mitigate the influence of age over the results, we first selected a subset of all subjects for whom ages were above $16$ years. Next, we converted their WRAT scores to $z$-scores, and only kept subjects whose absolute values of $z$-scores were above $0.5$. The low IQ group consisted of the subjects with $z$-scores smaller than $-0.5$, and the high IQ group consisted of the subjects with with $z$-scores larger than $0.5$. As a consequence, we were left with $n=224$ subjects that were separated into two groups according to IQ levels: the low and high IQ groups (Table \ref{t11}).

\begin{table}[!h]
\centering
\renewcommand\arraystretch{1.45}
\caption{Characteristics of the subjects in this study. SD: standard deviation.}
\label{t11}
\footnotesize
\setlength{\tabcolsep}{1mm}{
\begin{tabular}{cccc}
\hline
\multicolumn{1}{c}{Group}  & Age ($\text{Mean}\pm\text{SD}$) & Male/Female & WRAT score ($\text{Mean}\pm\text{SD}$) \\ \hline
Low IQ                 & $17.96\pm1.36$      & $31/59$       & $46.96\pm4.91$            \\
High IQ                & $18.54\pm1.50$      & $61/73$       & $64.25\pm2.57$              \\ \hline
\end{tabular}}
\end{table}

\begin{figure*}[!b]
  \centering
 \includegraphics[width=1.7\columnwidth]{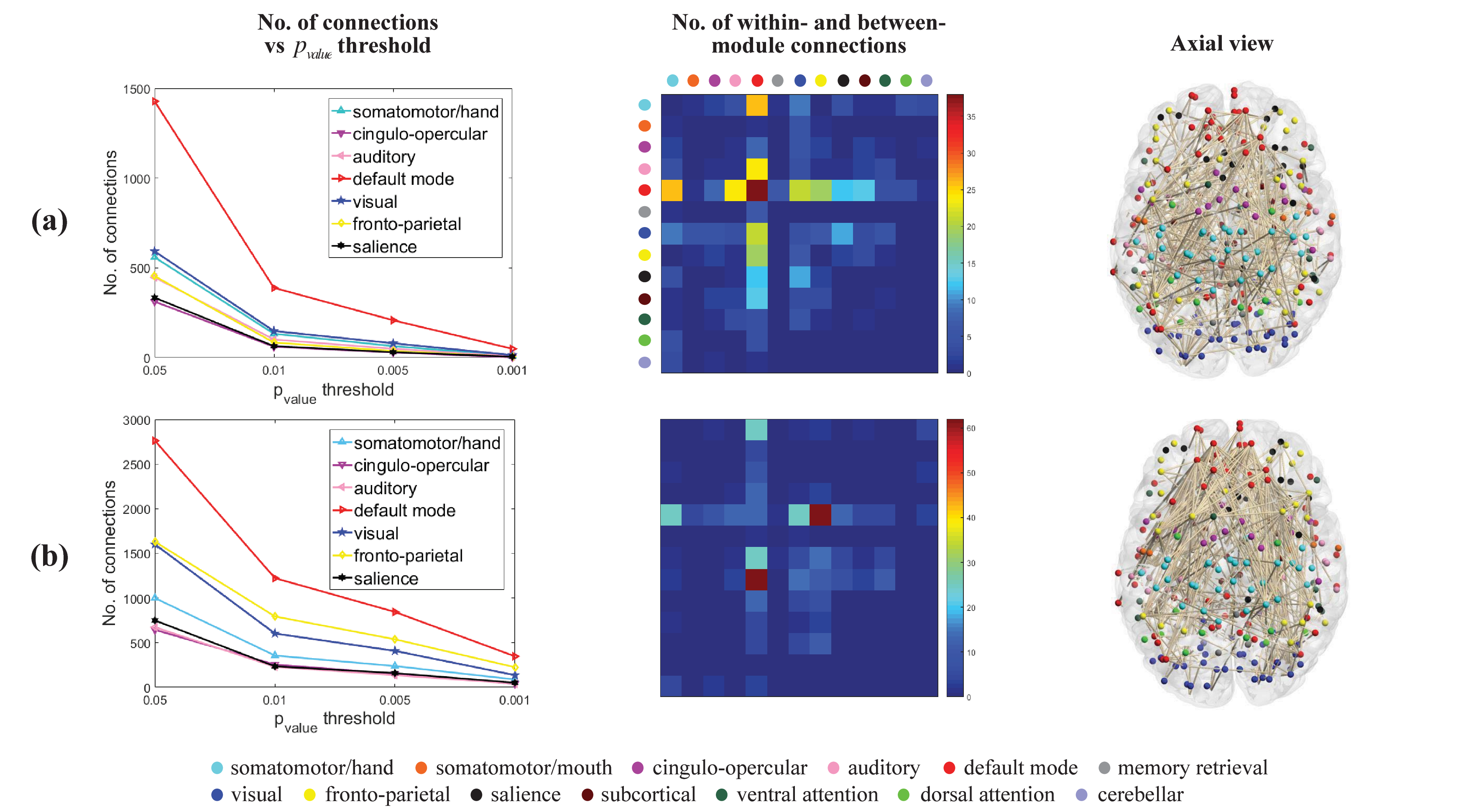}\\
  \caption{The brain FCN organizations associated with the connections differ significantly between the low and high IQ groups during (a) resting state and (b) fractal $n$-back task, respectively. The first column shows the number of connections by setting different $p_{\text{value}}$ thresholds. The last two columns display top $1\%$ connections. The second column shows the number of within- and between-module connections, and the third column shows three-dimensional axial brain views of the functional graph in anatomical space, where node colors indicate module membership.}\label{connect}
  \vspace{-0.1cm}
\end{figure*}

MRI examinations were conducted on a single $3$T Siemens TIM Trio whole-body scanner. Both task-based and resting-state images were collected using a single-shot, interleaved multi-slice, gradient-echo, echo planar imaging sequence. All the images were preprocessed in SPM12 (\url{www.fil.ion.ucl.ac.uk/spm/}), including motion correction, spatial normalization to standard MNI space, and spatial smoothing with a $3$mm FWHM Gaussian kernel. A regression procedure was applied to address motion-related influences and a $0.01$Hz--$0.1$Hz band-pass filter was applied to the functional time series. In resting-state, subjects were instructed to stay awake with the eyes open, fixate on the displayed crosshair, and keep still. In fractal $n$-back task to probe working memory, subjects were required to respond to a presented fractal only when it was the same as the one presented on a previous trial.
Based on a recently validated $264$-region functional parcellation scheme \cite{atlas}, $264$ ROIs were defined to describe the whole brain as $10$mm diameter spheres centered upon ROI coordinates. Thus, for each subject, each type of fMRI data can be represented by a matrix of which the rows correspond to the ROIs and the columns the time points.
All the fMRI data were centered and normalized by subtracting from each row the mean and dividing it by its standard deviation. We finally obtained two fMRI datasets, i.e., resting-state and fractal $n$-back task fMRI.

\subsubsection{Visualization of brain FCNs}
Recall that for each subject the FCN is defined by a $p\times p$ SPD matrix obtained in Subsection \ref{brain}, where $p=264$ is the number of ROIs. Within this network, there are $34716$ unique edges (or connections) and $14$ functional modules, i.e., somatomotor/hand, somatomotor/mouth, cingulo-opercular control, auditory, default mode, memory retrieval, visual, fronto-parietal control, salience, subcortical, ventral attention, dorsal attention, cerebellar, and uncertain. We sought to interrogate significantly different connections between low and high IQ groups. Two-sample t-tests were performed for each of the $34716$ Fisher z-transformed connection strength values in the network. In the first column of Fig. \ref{connect}, we displayed the number of connections by setting different $p_{\text{value}}$ thresholds (i.e., $0.05, 0.01,0.005,0.001$) in terms of $7$ typical modules. For ease of visualization, we ranked all connections according to their $t$-values, and selected the top $1\%$ of the connections (i.e., uncorrected, $p_{\text{value}}<5.21\times10^{-3}$ for resting-state, and $p_{\text{value}}<3.25\times10^{-4}$ for $n$-back task). The number of these selected connections differing between groups was assessed for each of the $13$ modules both for within- and between-module connections shown in the second column of Fig. \ref{connect}, and the corresponding three-dimensional axial views in anatomical space are visualized using the BrainNet Viewer \cite{xia}.
One can see that a majority of the significantly different connections associated with IQ were involved with the default mode, fronto-parietal control, and visual modules, which is in agreement with the reports in previous studies \cite{wholebrain1,wholebrain2,wholebrain3}. The default mode module has been linked to self-referential thought and autobiographical memory. The fronto-parietal module, including portions of the lateral prefrontal cortex and posterior parietal cortex, is thought to serve cognitive control abilities and working memory, among others. The visual module is related to the ability to process visual stimuli and to understand spatial relation between objects.

\subsubsection{Classification results}\label{claaaa}
We first assessed the classification performance for high vs low IQ when only one single dataset (resting-state FCNs or $n$-back task FCNs) was used with and without applying the DM. Second, we evaluated the classification performance when both resting-state and $n$-back task FCNs were used with applying the ADM. Third, we compared the performance of the proposed ADM based framework with that of several other common data fusion methods.

In nonlinear dimensionality reduction of the FCNs by the DM and the ADM, two important parameters have to be set, i.e., the kernel bandwidth $\sigma$ in the Gaussian kernel matrix and the target dimension of the reduced space, both of which influence the embedding and thus the subsequent classification results. Too small $\sigma$ will result in a sparse (or even disconnected) graph that is unable to capture the local structures around the data points, whereas too large $\sigma$ will cause a dense graph that may generate a redundant description of the data. Analogously, if the target dimension ($d$ in the DM or $\tilde{d}$ in the ADM) is too large or too small, the mapping will tend to be noisy and unstable or may not capture sufficient information about the manifold geometry.
Choosing parameters from a reasonable range is of importance. Notably, a max-min scheme has been suggested in \cite{xxxhhh} for choosing $\sigma$:
\begin{equation}\label{dingyi}
\sigma=C\cdot \max_{j}\min_{i,i\neq j}({\rm{d}}^2(\bm{R}_i,\bm{R}_j)),
\end{equation}
where $C$ is typically set in the range $[2,3]$. In this paper, we fixed $C=2$ for the kernel bandwidth in the DM. However, in the ADM, the unified kernel matrix (\ref{unif}) involves the product of two single kernel matrices. This insight indicates that the max-min measure for kernel bandwidth in the DM could be relaxed in the ADM. That is, smaller values for $C$ could be used to set $\sigma_1$ and $\sigma_2$ in (\ref{gauker}). Although an automated method for determining $\sigma_1$ and $\sigma_2$ has been proposed \cite{amdapp2}, we choose to tune them by cross-validation in this study.
Different values of the kernel bandwidth employed in our experiments were tested by setting $C\in\left\{0.2,0.4,\cdots,2\right\}$
for each dataset in the ADM. In both the DM and the ADM, the target dimension varied in the range of $\left\{10,20,\cdots,100\right\}$.

A $5$-fold cross-validation (CV) procedure was implemented to evaluate the classification performance in all experiments. The whole data were randomly partitioned into $5$ equal-sized disjoint subsets with similar class distributions. Each subset in turn was used as the test set and the remaining $4$ subsets were used to train the SVM classifier. Specifically, for every pair of training and test sets, the low-dimensional embeddings of the training set were first computed, and an SVM classifier was trained by the labeled samples in the embedded training set. Then, the low-dimensional embeddings of the test set were obtained by using the out-of-sample extension, and the trained SVM was applied to predict class labels of the samples in the embedded test set.
The classifier accuracy was estimated by comparing against the ground-truth labels on the test set. The test result in the CV was the average of the $5$ individual accuracy measures. The whole process was repeated $20$ times to reduce the effect of sampling bias, and the average classification accuracy (ACC) was computed over all $20$ realizations. All free
parameters, i.e., the kernel bandwidth and the target dimension, were tuned from their respective ranges by $5$-fold inner CV on the training set, and the parameters with the best performance in the inner CV were used in the testing.

\textit{3.1) Results of the DM and the ADM:}
The performance of the DM incorporating different distances (i.e., LEU (\ref{logeuc}), CK (\ref{chock}), and EU (\ref{euclidean})) on SPD matrices was tested for each single dataset of FCNs, respectively. To see if more or less significant information got lost after the embedding, we also vectorized the original high-dimensional FC data without applying the DM and then directly used them to train an SVM classifier. The results are reported in Table \ref{bar_new}.
We found that the classification performance using $n$-back task FCNs was usually better than that using resting-state FCNs. This highlights the importance of $n$-back task FCs in IQ classification. Compared with the results of the vectorized method, DM+CK and DM+EU got similar or even worse results, while DM+LEU made significant improvement. Among all the methods, DM+LEU achieved the best performance (DM+LEU vs the other methods: $p_{\text{value}}<0.0001$ for resting-state and $p_{\text{value}}<0.005$ for $n$-back task). It means that the incorporation of the LEU into the DM successfully extracted the most informative low-dimensional embeddings, but the incorporation of the other distances (i.e., the CK and the EU) into the DM did not.

\begin{table}[!h]
\centering
\caption{The comparison of classification results ($\text{ACC}\pm\text{SD}\,\%$) based on single fMRI dataset with/without applying the DM.}
\label{bar_new}
\renewcommand\arraystretch{1.45}
\footnotesize
\setlength{\tabcolsep}{1.7mm}{
\begin{tabular}{ccccc}
\hline
              & Vectorized & DM+LEU & DM+CK & DM+EU \\ \hline
resting-state & $65.63\pm2.53$  & $70.06\pm1.95$  & $66.96\pm2.47$ & $63.84\pm2.22$ \\ \hline
n-back task   & $69.29\pm1.91$  & $73.22\pm2.10$  & $70.76\pm1.97$ & $68.44\pm2.46$ \\ \hline
\end{tabular}}
\end{table}

We next compared the performance of the ADM for fusion of the two datasets of FCNs (i.e., resting-state FCNs and n-back task FCNs), as shown in the last row of Table \ref{t1}. The performance using the ADM based data fusion was better than that using the DM on any single dataset. In particular, ADM+LEU achieved $75.15\%$ classification accuracy, which was better than the results of the DM on any single dataset in Table \ref{bar_new} (e.g., in DM+LEU, $p_\text{value}=0.002$ for $n$-back task and $p_\text{value}<0.0001$ for resting-state), and made improvement of about $5\%$ in comparison to the vectorized method for each single dataset. It demonstrates the power of ADM based data fusion and also justifies the assumption that a proper fusion of different datasets can produce more coherent information useful to understand the observed phenomenon.
In accordance with the performance of both the DM and the ADM with respect to different distances on SPD matrices, the LEU always achieved the best result, the CK followed, and the EU was the lowest. This again indicates that it is important to consider the manifold property of SPD matrices to obtain low-dimensional embeddings, resulting in discrimination of individuals with different IQ levels.

\begin{table}[!h]
\centering
\renewcommand\arraystretch{1.45}
\caption{The comparison of classification results ($\text{ACC}\pm\text{SD}\,\%$) based on two fMRI datasets with applying different fusion methods.}\label{t1}
\footnotesize
\setlength{\tabcolsep}{1.7mm}{
\begin{tabular}{cccc}
\hline
Method                & LEU   & CK    & EU    \\ \hline
Concatenated DM \uppercase\expandafter{\romannumeral2}    & $73.89\pm2.56$ & $71.13\pm2.42$ & $67.74\pm2.43$ \\
Kernel-sum DM         & $74.55\pm2.24$ & $72.12\pm1.80$ & $69.55\pm2.50$ \\
Kernel-dot-product DM & $74.12\pm1.97$ & $71.07\pm1.46$ & $69.84\pm1.81$ \\
ADM                   & $75.15\pm1.72$ & $72.38\pm1.96$ & $69.58\pm2.09$ \\ \hline
\end{tabular}}
\end{table}

\begin{figure}[htbp]
\centering
\subfigure{
\begin{minipage}{0.22\columnwidth}
\centering
\hspace*{-2.5cm}\includegraphics[scale=0.235]{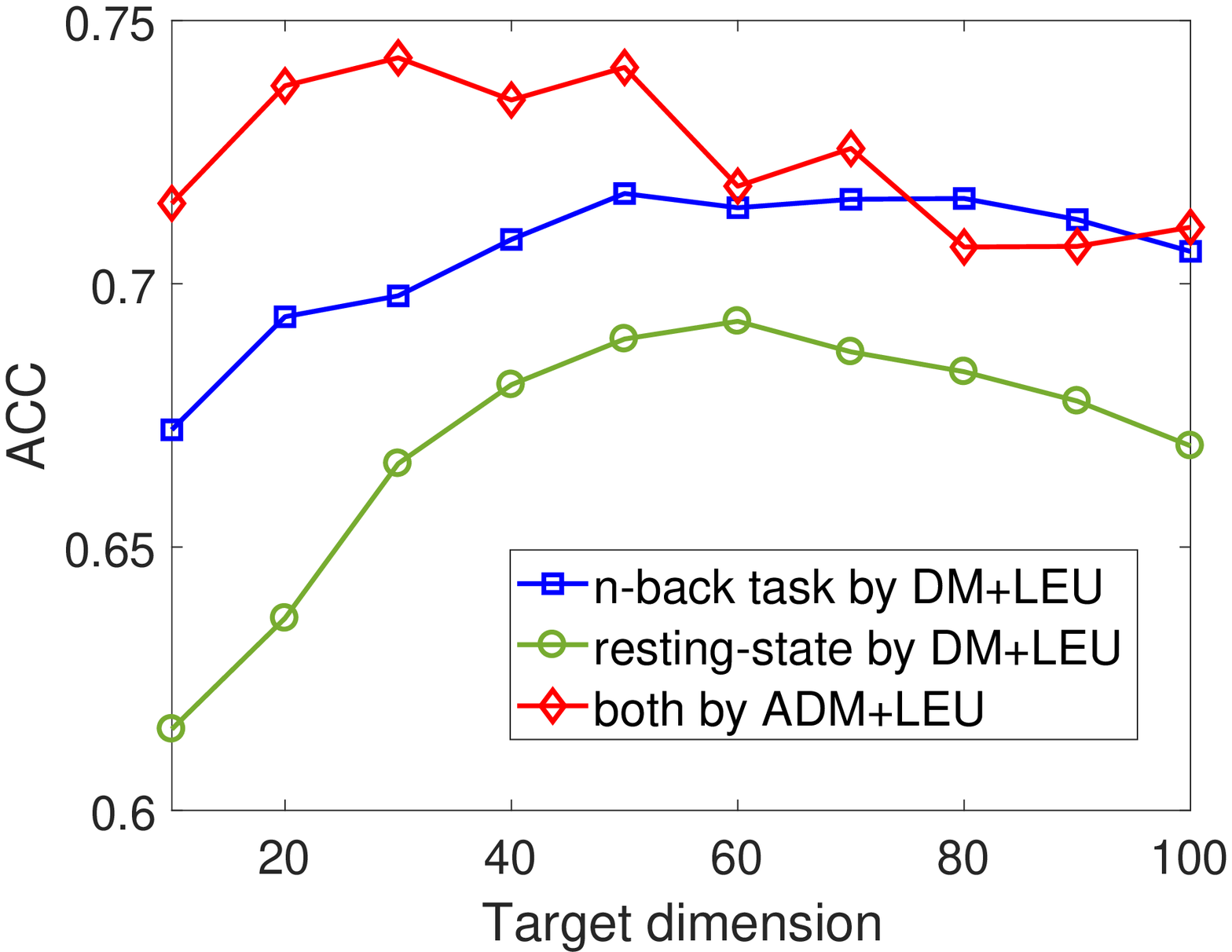}
\end{minipage}
}
\subfigure{
\hspace*{-0.25cm}\begin{minipage}{0.22\columnwidth}
\centering
\includegraphics[scale=0.235]{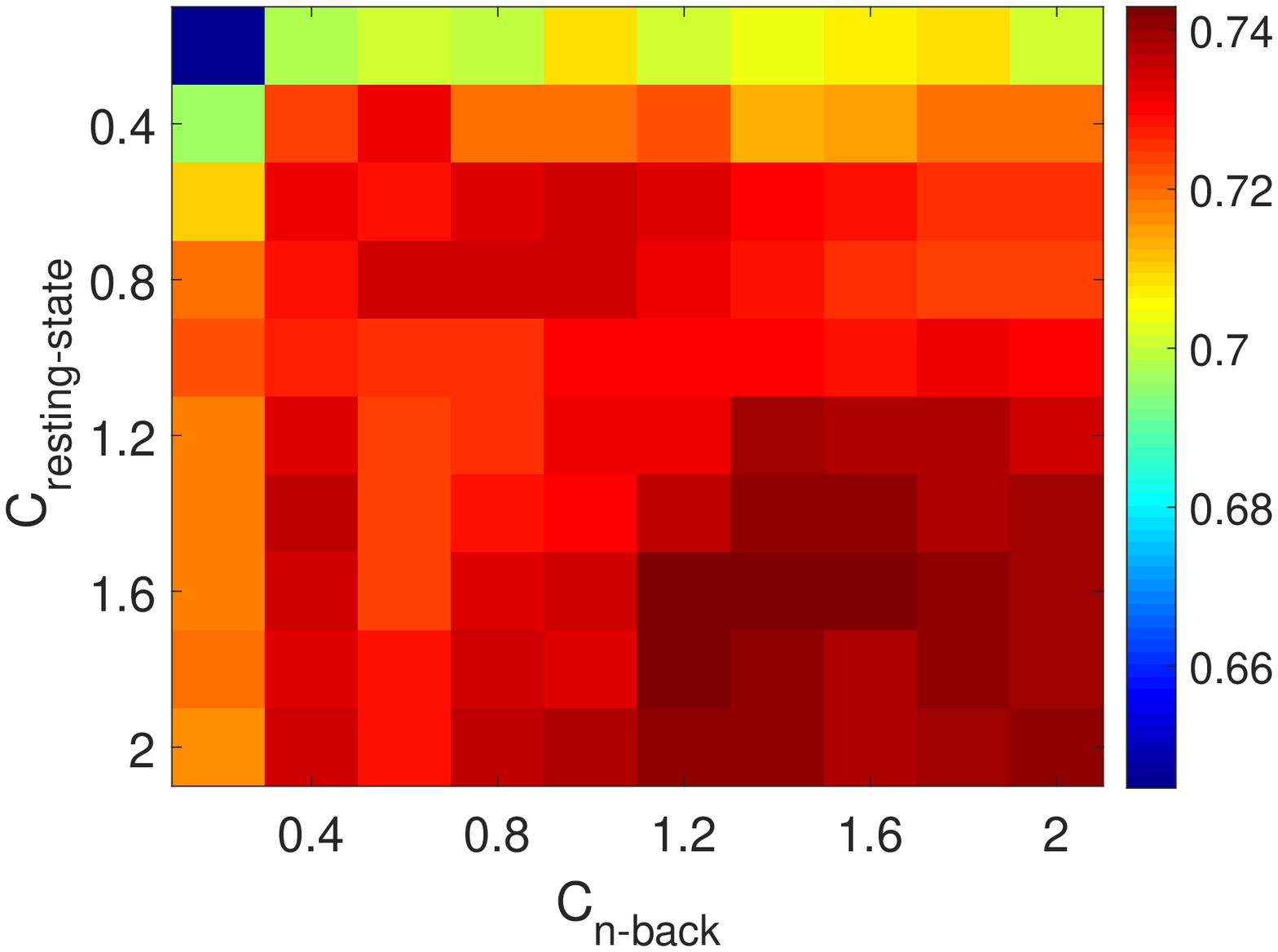}
\end{minipage}
}
\caption{The effect of parameters on the classification accuracy.}
\label{parameter}
\vspace{-0.1cm}
\end{figure}

We also investigated the effect of free parameters on classification performance in such a way that the parameters of interest were successively set to one combination across their ranges and for every setting of the parameters the testing accuracy was computed in the CV with the left-out parameters being optimally tuned. In the left of Fig. \ref{parameter}, the classification accuracies of DM+LEU for each single dataset and ADM+LEU for fusion of two datasets are shown with varying settings of the target dimension.
As seen from the figure, the target dimension has an important impact on the classification. If the selected target dimension is too small, the mapping will lose some important information. If the selected target dimension is too large, the embeddings will be still noisy and redundant such that they cannot effectively reflect the intrinsic structures of the original high-dimensional data. Both of the above cases will lead to poor classification accuracy.
Similarly, selecting optimum kernel bandwidths in the ADM plays a role in the classification performance. As shown in the right of Fig. \ref{parameter}, the parameters' sensitivity by changing values of $C_{resting-state}$ and $C_{n-back}$ in ADM+LEU is presented. We observed that the best parameter combination was always found in our experiments; for example, in ADM+LEU the selected target dimension was usually in the range of $[20,50]$, and the selected $C_{resting-state}$ and $C_{n-back}$ were usually in the range of $[1.2,1.8]$.

\textit{3.2) Comparison with other data fusion methods:}
To further demonstrate the strength of the ADM, we compare it with other data fusion methods described below.

\begin{itemize}[leftmargin=1.2em]
  \item[\romannumeral1] Concatenated DM \uppercase\expandafter{\romannumeral1}: concatenate all the features from two datasets into a single feature vector, and then apply the DM.
  \item[\romannumeral2] Concatenated DM \uppercase\expandafter{\romannumeral2} \cite{xxxhhh}: apply the DM to obtain low-dimensional embeddings of each dataset separately, and then concatenate the embeddings into a unified vector.
  \item[\romannumeral3] Kernel-sum DM \cite{kersum}: add up the similarity matrices constructed from each dataset to get a unified similarity matrix as $\bm{W}=\bm{W}^{(1)}+\bm{W}^{(2)}$, and then perform the rest of the procedures of the DM based on $\bm{W}$.
  \item[\romannumeral4] Kernel-dot-product DM \cite{DM1}: multiply the similarities matrices constructed from each dataset element by element to get a unified similarity matrix as $\bm{W}=\bm{W}^{(1)}\circ\bm{W}^{(2)}$, and then perform the rest procedures of the DM based on $\bm{W}$.
\end{itemize}

For fair comparison, all experiments for the above methods were implemented by the same evaluation framework as the ADM.
It turns out that the ADM with the LEU still achieved the highest accuracies among all the methods with all different distances on SPD matrices. It demonstrates the effectiveness of applying the LEU to measure the similarities of FCNs (e.g., compared with ADM+LEU, $p_{\text{value}}<0.0001$ for both ADM+CK and ADM+EU). When using the LEU on SPD matrices, the ADM performed improved results compared with the kernel-dot-product DM and the concatenated DM \uppercase\expandafter{\romannumeral2} with $p_{\text{value}}<0.05$, and yielded results most similar to the kernel-sum DM. For the EU and the CK, there were no substantial differences of accuracy between the kernel-sum DM, the kernel-dot-product DM, and the ADM. Importantly, similar to the ADM, the kernel-sum and kernel-dot-product DM methods define a unified similarity matrix that sums or multiplies the pairwise similarities between subjects from each dataset, resulting in a better combination of complementary information from each dataset. It is shown from Table \ref{t1} that both of them achieved better classification results than those using the concatenated methods (i.e., the concatenated DM \uppercase\expandafter{\romannumeral1} and the concatenated DM \uppercase\expandafter{\romannumeral2}).
In the concatenated DM \uppercase\expandafter{\romannumeral1}, the classification accuracy was only $69.64\%$. The classification performance of the concatenated DM \uppercase\expandafter{\romannumeral2} was slightly better than that of the concatenated DM \uppercase\expandafter{\romannumeral1}.
The poor classification performance based on the concatenated feature set in the concatenated methods may be largely ascribed to ignoring the mutual relations that exist between the datasets. This suggests that it is better to fuse heterogeneous datasets using kernel/similarity matrices rather than direct fusion in the original feature space.

\section{Discussion}\label{future}
\subsection{Most discriminative brain FCs}
It can be seen that the ADM with the LEU achieved the best classification performance. Equivalently, the low-dimensional embeddings obtained by this method best characterized the underlying data structures associated with IQ variability. Therefore, the alternating diffusion distance, defined by the Euclidean distance in the low-dimensional embedding space, i.e., $\lVert\bm{z}_i-\bm{z}_j\rVert_{2}$ for each pair of subjects $i$ and $j$, can provide a measure between subjects in terms of the common latent variables of interest extracted from the two sets of FCNs.
Based on the alternating diffusion distance, we attempted to evaluate the discriminative power of the features (i.e., FCs) according to their Laplacian scores \cite{lscore} as follows.

In each CV of the ADM with the LEU, we first learnt the embeddings $\left\{\bm{z}_i\right\}_{i=1}^{n}$ corresponding to the highest classification accuracy on the training set.
We then constructed a $k$-nearest-neighbor graph with $n$ nodes. The $i$-th node corresponds to $\bm{z}_i$. If $\bm{z}_i$ is among $k$ nearest neighbors of $\bm{z}_j$ or $\bm{z}_j$ is among $k$ nearest neighbors of $\bm{z}_i$, we put the edge
\begin{equation}\label{xinxin}
S_{i,j}=\exp\left(-\frac{\lVert\bm{z}_i-\bm{z}_j\rVert^{2}_{2}}{\gamma}\right),
\end{equation}
with $\gamma$ being set as $2\max\limits_{j}\min\limits_{i,i\neq j}(\lVert\bm{z}_i-\bm{z}_j\rVert^{2}_{2})$; otherwise, set $S_{i,j}=0$. This graph structure can nicely reflect the common manifold geometry of the data. Thus, the importance of a feature can be regarded as the degree to which the feature is consistent with the graph structure induced from (\ref{xinxin}).

\begin{figure}[!t]
  \centering
  \hspace*{-0.8cm}\includegraphics[width=1.2\columnwidth]{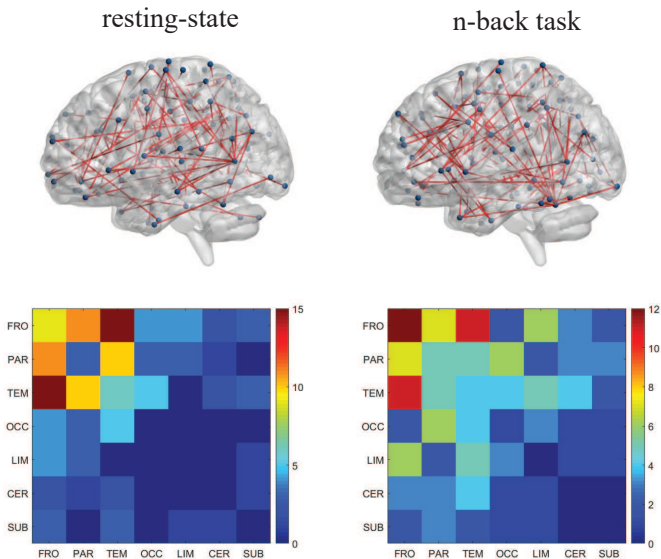}\\
  \caption{The visualization of the most discriminative $100$ FCs for resting-state and $n$-back task FCNs, respectively. The upper are brain plots of functional graphs in anatomical space, where the selected FCs are represented as edges. The lower are matrix plots, where the rows and columns represent the cortical lobes: frontal (FRO), parietal (PAR), temporal (TEM), occipital (OCC), limbic (LIM), cerebellum (CER), and sub-lobar (SUB).}\label{last}
  \vspace{-0.1cm}
\end{figure}

Let $f^{\rm{rs}}_{mi}$ denote the $m$-th resting-state FC of the $i$-th subject, and $\bm{f}^{\rm{rs}}_{m}=[f^{\rm{rs}}_{m1},f^{\rm{rs}}_{m2},\cdots,f^{\rm{rs}}_{mn}]^{T}$ with $n$ subjects. The Laplacian score of the $m$-th resting-state FC is defined by
\begin{equation}
L^{\rm{rs}}_{m}=\frac{\sum_{ij}(f^{\rm{rs}}_{mi}-f^{\rm{rs}}_{mj})^2S_{i,j}}{{\rm{Var}}(\bm{f}^{\rm{rs}}_{m})},
\end{equation}
where ${\rm{Var}}(\bm{f}^{\rm{rs}}_{m})$ is the estimated variance on the graph. By spectral graph theory, we compute ${\rm{Var}}(\bm{f}^{\rm{rs}}_{m})$ as
\begin{equation}
{\rm{Var}}(\bm{f}^{\rm{rs}}_{m})=\sum_{i}(f^{\rm{rs}}_{mi}-\mu^{\rm{rs}}_m)^2V_i,
\end{equation}
where $\mu^{\rm{rs}}_m=\sum_i\left(f^{\rm{rs}}_{mi}\frac{V_i}{\sum_{j}V_j}\right)$ and $V_i=\sum_{j}S_{i,j}$. Similarly, the Laplacian score $L^{\rm{nb}}_{m}$ of the $m$-th $n$-back task FC is also defined. Obviously, the smaller the Laplacian score is, the better the feature is. Since the Laplacian score of a feature is different in each CV, we averaged the Laplacian scores of each feature in all CV folds, and ranked the features according to their averaged Laplacian scores in increasing order.
We visualized $100$ resting-state and $n$-back task FCs with the smallest averaged Laplacian scores in Fig. \ref{last}, respectively, where $k$ was set as $10$. It is found that the majority of the selected FCs are located in frontal, parietal, temporal, and occipital lobes.

\subsection{Future work and limitations}
The free parameter tuning in the manifold learning methods, e.g., the kernel bandwidth and target dimension in the DM and the ADM in this paper, is crucial for classification. How to choose the optimal values for the free parameters remains an open and actively researched question. Although algorithms for automatic tuning of the optimal kernel bandwidth and target dimension in the DM and the ADM have been proposed in \cite{amdapp2}, they have been experimentally shown to be unsuitable for the datasets in this study. Therefore, in this paper we implemented grid search CV for parameter tuning. Note that Dudoit and van der Laan \cite{cross} have provided the asymptotic proof for choosing the tuning parameter with minimal CV error, which gives a theoretical basis for this approach.

In fMRI data analysis, not all ROIs are related to IQ differences. The ROIs are filtered to extract only those that can help to discriminate between high and low IQ. Therefore, feature selection could be performed to extract the most informative ROIs prior to constructing FCNs in our proposed framework. We will investigate the effect of using different feature selection approaches on the classification performance in future work.

In line with recent studies \cite{greene}, task fMRI data have a better prediction of IQ than resting-state fMRI data. Furthermore, it has been shown in \cite{gaosiyuan} that combining multiple different task fMRI datasets can significantly improve IQ predictive power, compared with using any single task fMRI dataset.
Apart from resting-state and $n$-back task fMRI datasets, there exist emotional task fMRI and single nucleotide polymorphism (SNP) datasets in the PNC. Therefore, an interesting future work is to fuse all the three neuroimaging datasets and one genomic dataset together by means of the ADM, which could capture more discriminative information and further improve the IQ classification performance.

\section{Conclusion}\label{condul}
In this paper, we considered a manifold based data fusion method (i.e., the ADM), by which the information from two datasets acquired by different sensors is diffused to extract the common information driving the phenomenon of interest, and simultaneously to reduce the sensor-specific nuisance. We tested the potential of the ADM for predicting IQ with the PNC dataset, resulting from a comprehensive study of brain development. Specifically, for each of resting-state and $n$-back task
fMRI, we first represented the FCN by a SPD matrix using the graphical LASSO for each subject. This results in two FCNs (or two SPD matrices), i.e., resting-state and $n$-back task FCNs, for each subject.
We next utilized the ADM to fuse the resting-state and $n$-back task FCNs to extract a meaningful low-dimensional representation. The obtained low-dimensional embeddings were used to train a linear kernel SVM classifier. The experimental results show that the prediction accuracy of the fused data by means of the ADM is larger than that of using any single set of FCNs, and the ADM also achieves superior classification performance in comparison with several other data fusion methods. Moreover,
in the construction of similarity matrices, we employed the Log-Euclidean manifold based metric to measure the distance between SPD matrices.  The effectiveness of incorporating it into the DM or the ADM was verified by the comparative experiments with the Cholesky metric and the traditional Euclidean metric on SPD matrices.

\section*{Acknowledgment}
The authors wish to thank the NIH (R01GM109068, R01MH104680, R01MH107354, R01AR059781, R01EB006841, R01EB005846, P20GM103472), and NSF (\#1539067) for their partial support.

\end{document}